\documentclass[letterpaper]{article} 
\usepackage[]{aaai2026}  
\usepackage{times}  
\usepackage{helvet}  
\usepackage{courier}  
\usepackage[hyphens]{url}  
\usepackage{graphicx} 
\usepackage{natbib}  
\usepackage{caption} 
\usepackage{bm} 
\usepackage{float} 
\usepackage{amsmath}
\usepackage{amsfonts}

\usepackage{newfloat}
\usepackage{listings}
\DeclareCaptionStyle{ruled}{labelfont=normalfont,labelsep=colon,strut=off} 
\floatstyle{ruled}
\newfloat{listing}{tb}{lst}{}
\floatname{listing}{Listing}
\title{A Network of Biologically Inspired Rectified Spectral Units (ReSUs) Learns Hierarchical Features Without Error Backpropagation}
\author{
    Shanshan Qin\textsuperscript{\rm 1}\thanks{Present address: Institute of Natural Sciences, Shanghai Jiao Tong University, Shanghai 200240, China.}, Joshua L. Pughe-Sanford\textsuperscript{\rm 1}, Alexander Genkin\textsuperscript{\rm 1}, Pembe Gizem Ozdil\textsuperscript{\rm 1,2}, Philip Greengard\textsuperscript{\rm 1}, Anirvan M. Sengupta\textsuperscript{\rm 3,4}, Dmitri B. Chklovskii\textsuperscript{\rm 1,5}\thanks{Corresponding author.} 
}

\affiliations{
    \textsuperscript{\rm 1}Center for Computational Neuroscience, Flatiron Institute, Simons Foundation, New York, NY, USA\\
    \textsuperscript{\rm 2}EDRS Doctoral Program, Swiss Federal Institute of Technology Lausanne (EPFL), Switzerland\\
    \textsuperscript{\rm 3}Center for Computational Mathematics, Flatiron Institute, Simons Foundation, New York, NY, USA\\
    \textsuperscript{\rm 4} Physics Department, Rutgers University, New Brunswick, NJ, USA\\
    \textsuperscript{\rm 5} Neuroscience Institute, NYU Langone Medical Center, New York, NY, USA

ssqin@sjtu.edu.cn, \{jpughesanford,pgreengard,dchklovskii\}@flatironinstitute.org, \{alexander.genkin,pgizemozdil,anirvans.physics\}@gmail.com
}

\begin{document}
\maketitle
\begin{abstract}
We introduce a biologically inspired, multilayer neural architecture composed of Rectified Spectral Units (ReSUs). Each ReSU projects a recent window of its input history onto a canonical direction obtained via canonical correlation analysis (CCA) of previously observed past–future input pairs, and then rectifies either its positive or negative component. By encoding canonical directions in synaptic weights and temporal filters, ReSUs implement a local, self-supervised algorithm for progressively constructing increasingly complex features.

To evaluate both computational power and biological fidelity, we trained a two-layer ReSU network in a self-supervised regime on translating natural scenes. First-layer units, each driven by a single pixel, developed temporal filters resembling those of Drosophila post-photoreceptor neurons (L1/L2 and L3), including their empirically observed adaptation to signal-to-noise ratio (SNR). Second-layer units, which pooled spatially over the first layer, became direction-selective—analogous to T4 motion-detecting cells—with learned synaptic weight patterns approximating those derived from connectomic reconstructions. 

Together, these results suggest that ReSUs offer (i) a principled framework for modeling sensory circuits and (ii) a biologically grounded, backpropagation-free paradigm for constructing deep self-supervised neural networks.
\end{abstract}

\begin{links}
    \link{Code}{https://github.com/ShawnQin/ReSU.git}
    \link{Supplemental Material }{https://github.com/ShawnQin/ReSU.git}
\end{links}

\section{Introduction}
Modern deep learning systems outperform human experts on vision and language benchmarks \cite{he2015delving,brown2020language}, as well as strategic games \cite{silver2016mastering}. Yet, they remain conspicuously inferior to biological intelligence in several fundamental aspects. First, contemporary models lack true compositional reasoning and long horizon planning \cite{keysers2019measuring}, cognitive abilities that humans take for granted. Second, they are vastly less efficient: training state-of-the-art networks requires megawatt-hours of energy and billions of labeled examples \cite{strubell2020energy,patterson2021carbon}, whereas the human brain draws only $\approx 20$W \cite{laughlin2001energy} and learns largely from self-supervision. Third, today’s models are prone to hallucinations, brittle generalization, and limited motor planning, rarely observed in biological agents \cite{ji2023survey,recht2019imagenet,geirhos2020shortcut}. If we wish to make AI more powerful, reliable and efficient, understanding the root cause of such gaps is needed.

A potential and often overlooked source of these shortcomings is architectural: nearly all modern AI systems rely on rectified linear units (ReLUs), a mid‑20th‑century abstraction drawn loosely from early electrophysiology, and are trained by the error backpropagation algorithm \cite{rumelhart1986learning}. ReLUs are inherently static---they sum only simultaneously arriving inputs before rectifying---thereby discarding the rich temporal dynamics that characterize biological neurons. Error backpropagation requires giant labeled training datasets and nonlocal interactions
across layers for which no biological substrate has been found.

Decades of experiments in model organisms have now described with cellular precision physiological and anatomical aspects of hierarchical feature emergence. Such growing body of knowledge creates an opportunity to design a more biologically grounded neuronal model, potentially yielding artificial networks that transcend the limitations of current ReLU/backpropagation‑based architectures.

\section{Our Contribution}
We introduce a biologically motivated, multi-layer network of Rectified Spectral Units (ReSUs). Each ReSU projects a recent window of its input history onto a canonical direction of canonical correlation analysis (CCA) of previously observed past-future input pairs and then rectifies the positive or negative component, implementing a dynamic, potentially local self-supervised algorithm. 

\begin{figure}[hb] 
\centering
\includegraphics[width=\linewidth]{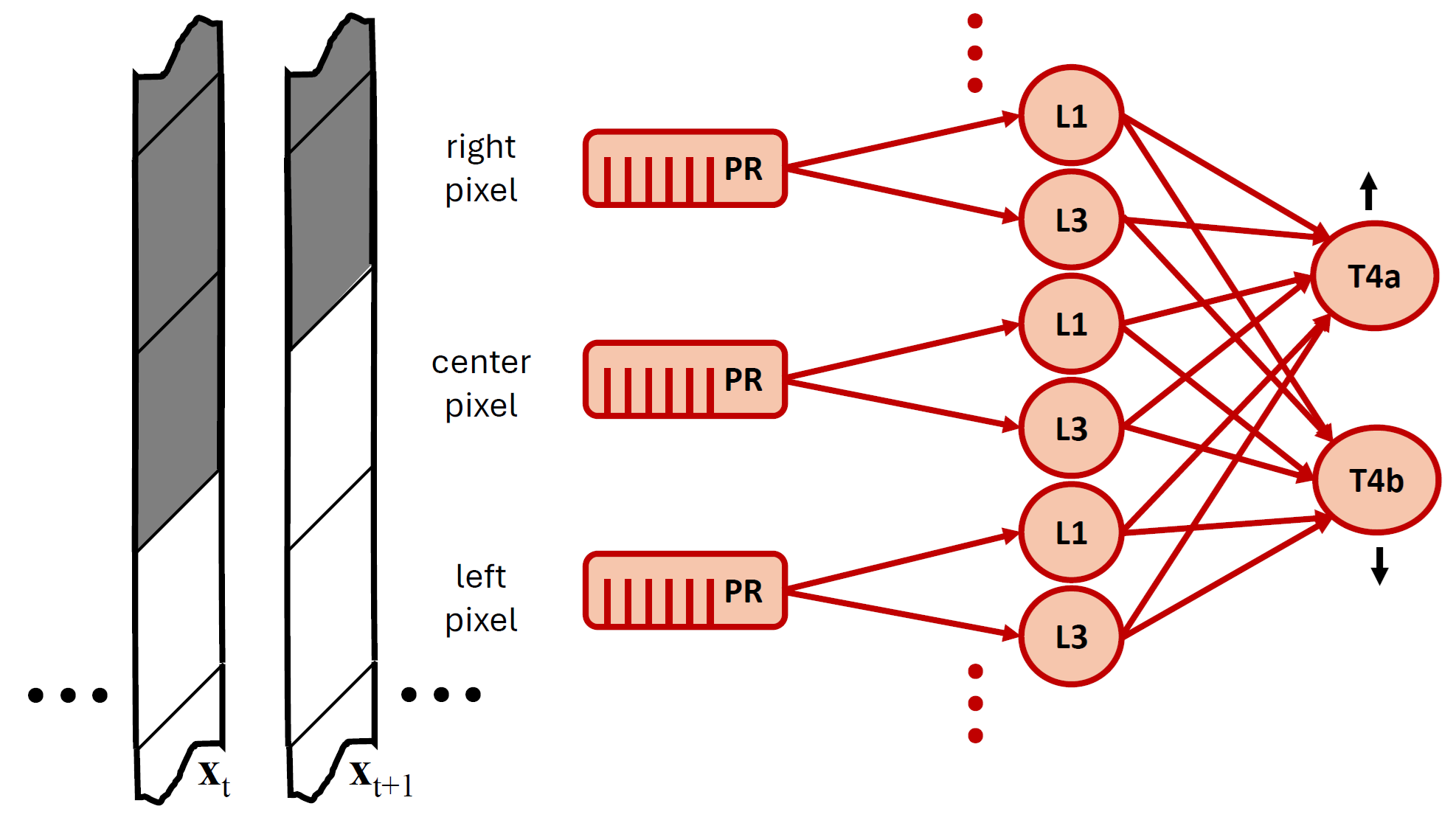}
\caption{{\it Drosophila}-inspired two-layer ReSU neural network trained in the self-supervised setting on translating natural images in 1D. Photoreceptors (PR) report the contrast levels of the corresponding pixels. Temporal filters of the first layer neurons (L1, L3) are computed by using CCA on past-future input sequences and act on PR outputs followed by rectification in L1. Second layer neurons (T4) pool information from 3 adjacent pixels. Past-future CCA of the outputs of the first layer learns a directionally selective  spatio-temporal filter for T4 (black arrows: preferred directions).} 
\label{fig:fig0}
\end{figure}

To evaluate both computational power and biological fidelity, we applied the ReSU framework to visual motion detection—a well-established, tractable, yet nontrivial benchmark task. The corresponding neural circuit in \textit{Drosophila} is exceptionally well characterized both anatomically and physiologically~\cite{takemura2013visual,takemura2017comprehensive,borst2023flies}, making it an ideal guide and testbed. We trained a \textit{Drosophila}-inspired two-layer ReSU network on natural-scene translations, Figure \ref{fig:fig0}, and obtained the following results:
\begin{itemize}
    \item Layer 1 units, each driven by a single pixel, learn temporal filters similar to those of the L1/L2 and L3 post-photoreceptor neurons of {\it Drosophila}, including their empirically observed adaptation to SNR.
    \item Layer 2 units pool outputs of Layer 1 across pixels and develop direction-selective responses analogous to T4 cells; their learned synaptic weight patterns approximate those found in connectomics reconstructions.
\end{itemize}

These findings demonstrate that ReSU networks:
(i) recover salient properties of biological circuits  offering a path to principled and interpretable modeling of hierarchical sensory processing,
(ii) generate hierarchical features offering a  back-prop-free path to constructing more brain-like deep artificial networks. Consequently, a biologically aligned ReSU architecture may inherit the advantages of the human brain over conventional ReLU/backpropagation-based architectures described in the Introduction.

\vspace{-2mm}
\section{Related Work}
Hierarchical features, like those observed in the cortex, are crucial for solving complex tasks. Yet despite decades of research, the computational principles underlying their emergence in the brain remain unknown.

Unsupervised single-layer networks derived from principled objectives using local (Hebbian and anti-Hebbian) rules and static units can learn principal components and sparse dictionaries \cite{oja1982simplified,olshausen1997sparse,hu2014hebbian,pehlevan2019neuroscience}. However, multiple attempts to “stack’’ such networks into deep architectures failed to generate richer hierarchical representations. Another principled approach is slow feature analysis (SFA) \cite{wiskott2002slow} which was implemented using local learning rules \cite{lipshutz2020biologically}. Whereas multi-layer SFA networks exist, obtaining cortex-like features with SFA requires first heuristically guessing hand-made nonlinear features, which is non-trivial.

Single-layer networks using static units and local (non-Hebbian) rules that learn CCA in unsupervised \cite{lipshutz2021biologically} and self-supervised \cite{golkar2020biologically} settings have been derived from principled objectives. However, because the processing in such networks is linear, unlike in the present work, ``stacking'' them into multi-layer networks would not produce complex features. Moreover, they do not capture the rich dynamics of biological neurons. 

In deep networks built on the predictive coding principle \cite{rao1999predictive,rao2024sensory,whittington2017approximation}neurons compute and output prediction errors rather than hierarchical features typically observed in the sensory pathways. Also, neuronal temporal properties in early sensory processing do not seem to align with the predictive coding architecture~\cite{druckmann2012mechanistic}.

Deep networks constructed using more complex local learning rules integrating self-supervised learning (such as contrastive learning), Hebbian plasticity, and predictive coding can learn hierarchical and invariant object representations \cite{illing2021local,dora2021deep,halvagal2023combination}. However, the inherent complexity of these multi-principle frameworks does not facilitate our understanding of the computational primitives operating at the single neuron level, making it difficult to isolate and identify the core mechanisms underlying biologically plausible learning.

Deep networks of static ReLUs trained by back-propagation produce hierarchies matching cortical organization \cite{yamins2014performance} yet rely on error signals or weight symmetries that are difficult to reconcile with neurobiology \cite{crick1989recent,lillicrap2020backpropagation}. Efforts to render backpropagation more biologically plausible---random feedback alignment \cite{lillicrap2016random,nokland2016direct}, dendritic error segregation \cite{guerguiev2017towards}, equilibrium propagation\cite{scellier2017equilibrium}--- result in performance degradation. 

Our approach is conceptually related to JEPA~\cite{LeCun2022PathAMI} and VAMPnets~\cite{mardt2018vampnets}, both of which learn representations predictive of future latent states without reconstructing raw inputs.
Whereas JEPA and VAMPnets rely on parameterized encoder–predictor networks trained using backprop end-to-end, ReSU derives analogous predictive primitives analytically via past–future CCA, yielding locally learnable, biologically interpretable units that can be hierarchically composed into deep networks.

A recent application of a backpropagation-trained network of dynamical units to the same {\it Drosophila} circuit~\cite{lappalainen2024connectome} produced a mechanism reminiscent of that observed experimentally. However, these networks are trained in a biologically implausible supervised setting using visual motion ground truth and do not consistently reproduce the experimentally observed motion detection computations. This limits their relevance for understanding biologically plausible learning of complex features, particularly in less well-characterized systems.

\section{Truncated CCA Maximizes Past-Future Mutual Information for OU Processes}
\label{sec:cca}

\begin{figure}[ht]
\centering
\includegraphics[width=\linewidth]{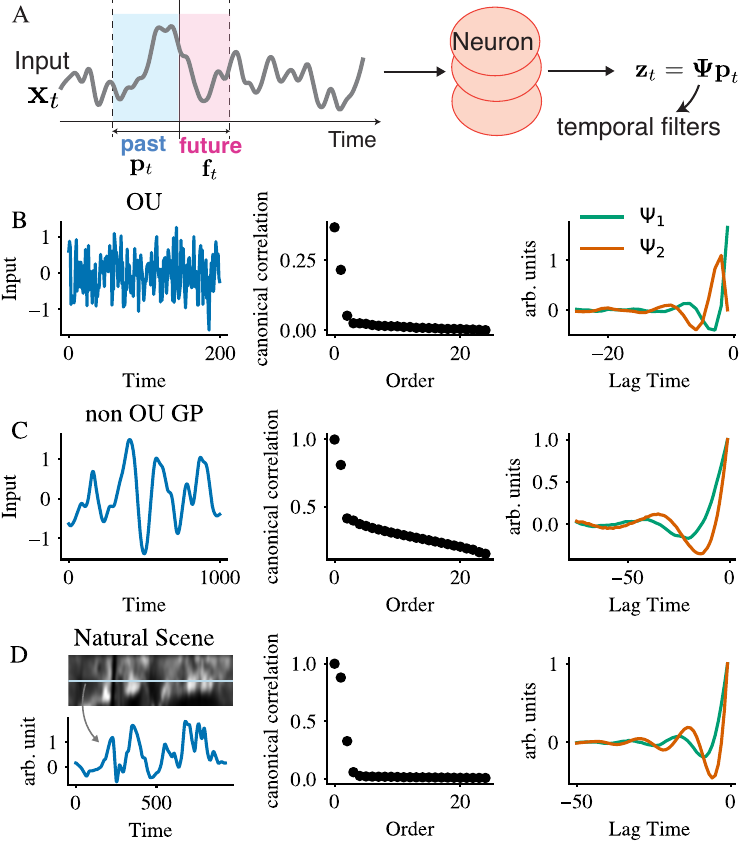}
\caption{Past–future CCA as a computational primitive applied to different data streams. (A) A sensory stimulus is represented by past and future lag vectors. Each neuron computes its output as a linear projection of the input onto a temporal filter learned from previous observations of the same stimulus. Results of CCA applied to (B) a two-dimensional OU process projected onto one dimension, (C) a non-OU Gaussian process with a rational quadratic kernel, and (D) a contrast profile obtained by scanning a natural image at constant velocity. The first column shows example time series, the second column displays the canonical correlations ($\sigma_i$), and the third column shows the first two normalized temporal filters (canonical directions). In all numerical simulations, a small Gaussian white observation noise was added to the input signal.}
\vspace{-2mm}
\label{fig:fig1}
\end{figure}

We start by postulating that sensory stimuli can be modeled as observations of stochastic dynamical systems. Then the goal of neurons is to learn such dynamics to support prediction and control. As a fully tractable first step towards learning the dynamics of natural stimuli, we consider a discrete-time, linear, time-invariant stochastic system, known as a partially observed multivariate Ornstein–Uhlenbeck (OU) process (Figure \ref{fig:fig1}B) and defined by :
\begin{align}
\label{dyn}
    {\bf x}_{t+1} &= {\bf A x}_{t} + {\bf Bv}_{t}, \\
    y_t &= {\bf Cx}_t + w_t,
    \label{dyn1}
\end{align}
where \({\bf x}_t \in \mathbb{R}^n\) is the latent state vector, \(y_t\in \mathbb{R}\) is the observed scalar time series, each componet of \({\bf v}_t \in \mathbb{R}^n\) and \(w_t\) are independent, identically distributed Gaussian noise sources with zero mean and unit covariance. \({\bf A, B, C}\) are constant but unknown matrices of compatible dimensions.

In this framework, the state vector \({\bf x}_t\) acts as an informational interface between past noise inputs, \({\bf v}_t\), and future observation, \(y_t\). To obtain a reduced-dimensional representation of \({\bf x}_t\), that preserves maximal mutual information with the future, we apply balanced truncation~\cite{katayama2005subspace}. Although \({\bf x}_t\) is not directly observable, it can be inferred from sequences of partial observations, \(y_t\), using past and future lag vectors, as illustrated in Figure \ref{fig:fig1}A:
\begin{equation}\label{pf_vects}
\begin{aligned}
    {\bf p}_t &= [y_t, y_{t-1}, \dots, y_{t-m+1}]^\top, \\
    {\bf f}_t &= [y_{t+1}, y_{t+2}, \dots, y_{t+h}]^\top,
\end{aligned}
\end{equation}
where \(m,h\ge n\) denote the lengths of the past (memory) and future (horizon) lag vectors, respectively. Although (centered) ${\bf p}_t$ and ${\bf f}_t$ are not minimal representations of latent state ${\bf x}_t$, they form sufficient statistics for reconstructing its dynamics and estimating its predictive subspace.

The low dimensional latent representation is obtained via a linear projection of the past, Figure \ref{fig:fig1}A: 
\begin{align}
\label{z}
\mathbf{z}_t = \mathbf{\Psi}\mathbf{p}_t, \quad {\bf z}_t \in \mathbb{R}^r,
\end{align}
where the matrix $\mathbf{\Psi}\!\in\!\mathbb{R}^{r\times m}$ compresses the past while retaining maximal information about the future.

For the OU process (Eq.\ref{dyn}), the information captured in ${\bf z}_t$ about the future can be quantified by the correlation between ${\bf z}_t$ and the whitened future input, ${\bf f}_t$, leading to the following constrained optimization problem \cite{arun1990balanced},
\begin{equation}
\label{eq:CCA}
\hspace{-2mm}
    \max_{\mathbf{\Psi}\in\mathbb{R}^{r\times m}}
    \bigl\lVert \mathbb{E}\!\bigl[\mathbf{C}_{ff}^{-1/2}\,\mathbf{f}_t(\mathbf{\Psi}\mathbf{p}_t)^\top\bigr]\bigr\rVert_{F},
    \;
    \text{s.t.}\;
\mathbf{\Psi}\mathbf{C}_{pp}\mathbf{\Psi}^\top=\mathbf{I}_r,
\end{equation}
where the covariance matrices are defined as ${\bf C}_{ff} = \mathbb{E}[\mathbf{f}_t\mathbf{f}_t^\top], {\bf C}_{fp} = \mathbb{E}[\mathbf{f}_t\mathbf{p}_t^\top], {\bf C}_{pp} = \mathbb{E}[\mathbf{p}_t\mathbf{p}_t^\top]$. Substituting $\widetilde{\mathbf{\Psi}}=\mathbf{\Psi}\mathbf{C}_{pp}^{1/2}$ into Eq. \ref{eq:CCA} yields the equivalent form:
\begin{equation}
\label{eq:CCA2}
 \hspace{-2mm} \max_{\widetilde{\mathbf{\Psi}}\in\mathbb{R}^{r\times m}}
    \bigl\lVert \mathbf{C}_{ff}^{-1/2}\,\mathbf{C}_{fp}\,\mathbf{C}_{pp}^{-1/2} \,\widetilde{\mathbf{\Psi}}^\top\bigr\rVert_{F},
    \;
    \text{s.t.}\;
\widetilde{\mathbf{\Psi}}\widetilde{\mathbf{\Psi}}^\top=\mathbf{I}_r,
\end{equation}
which is solved via singular value decomposition (SVD) of the whitened cross-covariance matrix $\mathbf{C}_{ff}^{-1/2}\mathbf{C}_{fp}\mathbf{C}_{pp}^{-1/2}
    =\mathbf{U\Sigma V}^\top$ yielding:
    \vspace{-2mm}
\begin{equation} \label{eq:filter}
    \mathbf{\Psi} = \mathbf{V}_r^\top\,\mathbf{C}_{pp}^{-1/2},
\end{equation}
where ${\bf V}_r$ consists of the $r$ right singular vectors corresponding to the largest singular values, $\sigma_1 \dots \sigma_r$, which form the diagonal of $\bf \Sigma$, Figure \ref{fig:fig1}B. We also define $\mathbf{\Phi} = \mathbf{U}_r^\top \mathbf{C}_{ff}^{-1/2}$
from the corresponding $r$ left singular vectors. In practice, inverses of covariances may require $l_2$-norm regularization.

This procedure is equivalent to performing CCA between $\mathbf{f}_t$ and $\mathbf{p}_t$, where the singular values $\sigma_i$ represent the canonical correlations~\cite{arun1990balanced,chechik2003information}. In other words, each $\sigma_i$ quantifies the correlation between the $i$-th pair of canonical variates, $(\mathbf{\Psi}\mathbf{p}_t)_i$ and $(\mathbf{\Phi}\mathbf{f}_t)_i$.

The mutual information between $\mathbf{z}_t$ and future $\mathbf{f}_t$ is then given by \cite{arun1990balanced,chechik2003information} (see Supplementary Material):
\vspace{-2mm}
\begin{equation}\label{mutual_info}
  I_r = -\tfrac{1}{2}\sum_{i=1}^{r}\log\!\bigl(1-\sigma_i^{2}\bigr).
  \vspace{-1mm}
\end{equation}
Thus, for an OU process, truncated CCA yields an optimal linear projection of the past lag-vector onto the $r$-dimensional subspace that maximizes the mutual information with the future, providing a principled approach to extracting predictive latent representations from sensory input. 

\section{Truncated CCA Maximizes Past-Future Mutual Information for Gaussian Processes}\label{sec:gps}
As the next step towards the dynamics of natural stimuli, we demonstrate that past-future CCA is optimal not just for OU but for a more general class of one-dimensional Gaussian processes (GPs)—specifically, discrete-time GPs with translation-invariant and integrable covariance kernels. 

GPs are flexible, non-parametric models widely used in machine learning and statistics, Figure \ref{fig:fig1}C. They are parameterized by covariance kernels that allow practitioners to specify features such as the rate of autocovariance decay in time and spectral domains. This flexibility enables us to generate synthetic data with features that mimic natural stimuli. 

Here, we assume that stimuli are generated by GPs defined on a set of equispaced points $t_1, t_2, \dots$ with stationary and integrable covariance 
kernel $k(t_i - t_j)$. 

Constructing the GP's past/future vectors $\mathbf{p}_t, \mathbf{f}_t$ of \eqref{pf_vects} to 
be lag vectors of length $\ell$, the $2\ell \times 2\ell$ covariance matrix $\bm \Sigma$ of 
$\mathbf{p}_t, \mathbf{f}_t$ for all $i, j$ is $\bm \Sigma_{ij} = k(t_i - t_j)$. Similarly, representing $\bm \Sigma$ as a $2 \times 2$ block matrix, we have 
\vspace{-1mm}
\begin{align}
\bm \Sigma = \begin{bmatrix}
\mathbf{C}_{pp} & \mathbf{C}_{pf} \\
\mathbf{C}_{fp} & \mathbf{C}_{ff}
\end{bmatrix}.
\end{align}
The canonical directions can then be computed as described in the previous section, Figure \ref{fig:fig1}C. For a given lag-vector length $\bm p_t$, CCA maximizes correlation and mutual information with the future. The necessary length of the past lag vector depends on the decay of the covariance kernel, or how far back in time one needs to go until including more past observations is no longer informative about the future. 

For most commonly-used GP kernels, the canonical correlations will decay rapidly \cite{ambikasaran2016gps}. Intuitively, this means that there's only so much information from the past that can be used to predict the future. For complicated kernels, such as oscillatory ones, the future may depend on many independent features of past observations. For these kernels, the canonical correlations would not decay rapidly (see Supplementary Material for further discussion).

\begin{figure}[ht] 
\centering
\includegraphics[width=\linewidth]{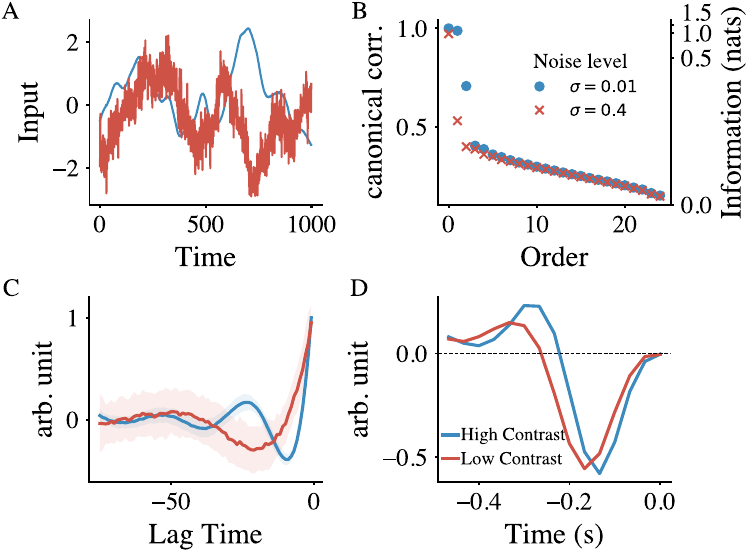}
\vspace{-2mm}
\caption{Dependence of past–future CCA results on observation noise in Gaussian processes. (A) Example time series generated from a Gaussian process with a rational quadratic kernel for high (blue) and low (red) SNR, varied through observation noise. (B) Correlation and mutual information between past projections onto canonical directions and the corresponding future signals. (C) Second canonical directions for high (blue) and low (red) SNRs. As observation noise increases, the filter shape transitions from multi-lobed to single-lobed. Shaded regions denote standard deviation across realizations. (D) Experimentally measured temporal filter of a retinal ganglion cell adapts to low contrast, which corresponds to lower SNR~\cite{liu2015spike}. These filters must pass through the origin (0,0), unlike the model filters (panel~C), because physiological filtering is constrained to be both causal and continuous—conditions not enforced in our model.}
\label{fig:fig2}
\vspace{-2mm}
\end{figure}

Next, we demonstrate the truncated CCA framework for GPs through specific examples, Figure \ref{fig:fig2}. We generate a synthetic dataset comprised of time series sampled from a GP with a rational quadratic kernel, i.e., $k(t,t') = \left(1+ \frac{|t-t'|^2}{2\alpha l^2}\right)^{-\alpha}$ for $\alpha, l > 0$. Here, $l=1$ controls the timescale and $\alpha=1$ the decay in the time domain of the autocovariance function. We choose this kernel because GPs with this kernel have properties that mimic statistics in natural images, where spatial correlations of contrast decay as a power law~\cite{ruderman1994statistics}. Then we derive the past temporal filter by solving the rank-2 CCA optimization problem \eqref{eq:CCA} with $m = 75, h = 50$. The resulting filter $\Psi$ has multiple lobes (blue line in Figure  \ref{fig:fig2}C). The latent representation $\bm z_t$ is obtained by projecting the past lag vectors $\mathbf{p}_t$ onto $\Psi$, which is the projection that maximizes mutual information with the future (Figure  \ref{fig:fig2}C).

Finally, we investigated how the temporal filter adapts to changes in stimulus statistics, particularly the SNR. Many sensory neurons exhibit such adaptation (Figure~\ref{fig:fig2}D)~\cite{srinivasan1982predictive,van1992theoretical,liu2015spike}, a hallmark of efficient and predictive coding~\cite{srinivasan1982predictive,van1992theoretical,chalk2018toward}. To test whether our model captures this property, we added Gaussian white noise of varying amplitudes to the observed time series and trained the model to derive the corresponding optimal filters. As shown in Figure~\ref{fig:fig2}C, under low noise, the filter exhibits multiple lobes, whereas with increasing noise it gradually transitions to a single-lobed shape—mirroring the adaptive temporal filters measured experimentally in sensory neurons~\cite{srinivasan1982predictive,van1992theoretical,liu2015spike}. Moreover, this adaptation can occur dynamically: when the SNR changes abruptly, the filter adjusts to the new input statistics within roughly ten lag-vector lengths (see Section 3 and Fig. S2 in the Supplementary Material). Thus, the adaptive properties of neuronal temporal filters in our model emerge naturally as a consequence of optimizing predictive information.

\section{ReSUs Trained on Natural Images Reproduce Physiological Responses}

In this Section we use the past-future CCA framework to model neurons by applying it to natural stimuli. As each neuron can only output a scalar as a function of time---a synaptic vesicle release rate or a firing rate---we need to specify how to partition the $r$-dimensional subspace, Eq. \ref{z}, among neurons. Gaussian information bottleneck does not specify this as any basis obtained by the orthogonal rotation of the singular vectors should be equally good \cite{chechik2003information}.  

We propose that a layer of $r$ neurons compresses the inputs by projecting $\mathbf{p}_t$ onto the temporal filters given by the rows of $ \mathbf{\Psi}$ yielding ${\bf z}_t$, Eq. \ref{z}. As the components of ${\bf z}_t$ are whitened, Eq. \ref{eq:CCA}, this has the following advantages. Firstly, it allows straightforward change of the rank in the optimal representation by adding and removing neurons in the order of the corresponding singular values without relearning temporal filters. Secondly, channel whitening is a desirable feature in case of noisy downstream communication and processing \cite{linsker1988self,plumbley1993hebbian}. 

We test this model by comparing its predictions with experimental data from the L1–L3 neurons in the \textit{Drosophila} visual system. These cells receive direct input from photoreceptors that sample the same location in visual space. Assuming that photoreceptor outputs approximate the local contrast of natural scenes, L1–L3 can be viewed as processing a scalar contrast time series for a single pixel, largely independent of neighboring pixels~\cite{rivera2014wiring,borst2023flies}. To model the visual input experienced during ego-motion, we simulate the eye scanning a natural scene at constant velocity by sampling pixel intensities along a straight line of a natural image (Figure~\ref{fig:fig1}D). The resulting contrast time series is used to compute neuronal filters via past–future CCA (Eq.~\ref{eq:filter}), which we then compare to the experimentally measured temporal response profiles of L1–L3~\cite{ketkar2022first}.

The first filter is essentially low-pass, while the second filter is a smoothed temporal derivative, predicting a similar output to the activation measured for L3 and L1-L2 respectively (Figure \ref{fig:fig3}, see Supplementary Material for details). 
To build intuition for our result, consider the so-called “dead leaves model” of natural images. This model assumes that the visual scene is composed of distinct objects, each with a fixed but different contrast from its neighbors. Scanning such a scene along a straight line therefore produces a time series of constant-contrast plateaus separated by transitions. Because in \textit{Drosophila} each photoreceptor samples a relatively large region of visual space ($\approx 5^\circ$), smaller than typical object boundaries, the shape of these contrast transitions reflects the optical properties of the eye and is likely to be stereotyped.
\begin{figure}[ht]\label{fig:fig3}
\centering
\includegraphics[width=\linewidth]     {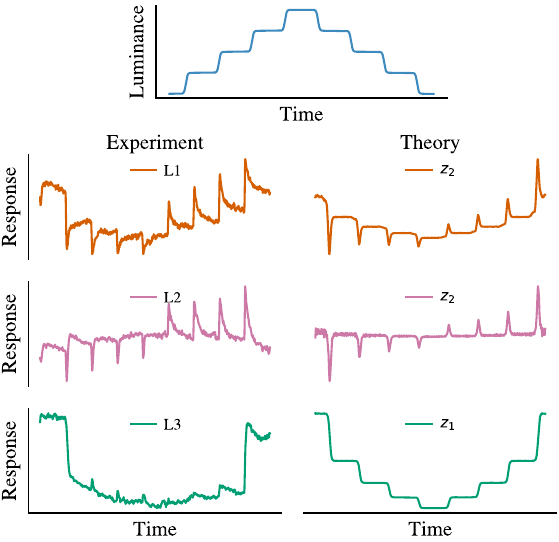}
\caption{Experimentally measured and theoretically predicted responses to the staircase stimulus. (Top) Luminance as a function of time. (Left column) Experimental measurements of average neuronal activity via calcium imaging in three post-photoreceptor neurons in \textit{Drosophila}: L1, L2 and L3 \cite{ketkar2022first}. (Right column) Output from the first and second temporal filters derived from past-future CCA of natural images. Orange and pink traces differ because of variation in the SNR, i.e., observation noise level: $\sigma = 0.5, 0.05$ respectively.}
\label{fig:fig3}
\vspace{-2mm}
\end{figure}

We can now interpret the roles of L1–L3 neurons in terms of maximizing the mutual information between past and future contrast time series. L3, which outputs a low-pass–filtered version of contrast, captures the strong correlations that persist during contrast plateaus. In contrast, L1 and L2 act as smoothed temporal derivative filters, emphasizing correlations in the rate of change during transitions. Because L1 outputs are rectified by their downstream synapses, they respond selectively to increases in contrast—an ON response. Conversely, L2 outputs are rectified in the opposite direction and respond to decreases in contrast—an OFF response. The plateaus and transitions can thus be viewed as “slow features” that exhibit temporal inertia and are therefore useful for prediction~\cite{wiskott2002slow}. These slow features can be further analyzed and combined downstream, as discussed in the next section.

The linear projection described above would provide an optimal representation under the information bottleneck principle~\cite{arun1990balanced,chechik2003information} if sensory stimuli were generated by linear time-invariant dynamics driven by white Gaussian noise, Eqs.~\eqref{dyn},\eqref{dyn1}. However, this conclusion conflicts with experimental observations showing pronounced output nonlinearities, such as rectification, in most neurons. While such nonlinearities may partly reflect metabolic efficiency~\cite{gjorgjieva2014benefits}, here we view them as computationally beneficial for extracting predictive latent variables from observations of \textit{nonlinear}—or equivalently, time-varying linear—dynamics. These dynamics can be approximated by switching linear systems, where switching occurs when neural activity crosses a rectification threshold. In this view, each neuron performs a soft clustering of dynamical trajectories and outputs a non-negative membership index~\cite{pehlevan2017clustering,pughe2025neurons}. This clustering perspective naturally arises within the Koopman operator framework for nonlinear dynamics which offers a linear representation in lifted feature spaces learnable from data and potentially applicable to hierarchical, deep architectures~\cite{mezic1999method,williams2015data,kaiser2021data,klus2024dynamical,pughe2025neurons}.

Motivated by the above considerations and the non-negativity of neuronal outputs, we propose the Rectified Spectral Unit (ReSU)—a neuron model that projects past input onto a canonical direction and rectifies the projection’s positive or negative component: 
\begin{equation}\label{ReSU}
    \begin{aligned}
    {\rm ON\;ReSU:}\quad 
    z^+_{t,i} &= \max\!\left[{\bf v}_i^\top {\bf C}_{pp}^{-1/2}{\bf p}_t,\, 0\right], \\
    {\rm OFF\;ReSU:}\quad 
    z^-_{t,i} &= \max\!\left[-{\bf v}_i^\top {\bf C}_{pp}^{-1/2}{\bf p}_t,\, 0\right],
\end{aligned}
\end{equation}
where ${\bf v}_i$ is the $i$-th singular vector and $i$-th column of ${\bf V}$. The term “Spectral” is used in its linear-algebraic sense and appears in the abbreviation because the projection is derived from the eigendecomposition of the whitened past–future covariance Gramian~\eqref{eq:CCA2}. Following biology, we model L3 using a non-rectified projection \eqref{z} with $i=1$, and L1, L2 using $i=2$ ON, OFF ReSUs \eqref{ReSU}, respectively.

Compared to ReLUs, whose synapses are learned by error backpropagation, ReSUs appear more biologically plausible. This is because supervised learning requires labeled ground truth data which is typically not available to neurons and backpropagation relies on nonlocal learning rules for which no biological substrate has been found. At the same time, ReSUs are self-supervised and do not require labeled ground truth. Since synaptic weights in ReSU networks are obtained by past-future CCA of the input, they don't require communication across layers as does backpropagation. Moreover, \cite{lipshutz2021biologically,golkar2020biologically} have shown that CCA (although not past-future) can be implemented by local learning rules (see Discussion).

\begin{figure*}[h]
  \centering
\includegraphics[width=\textwidth]{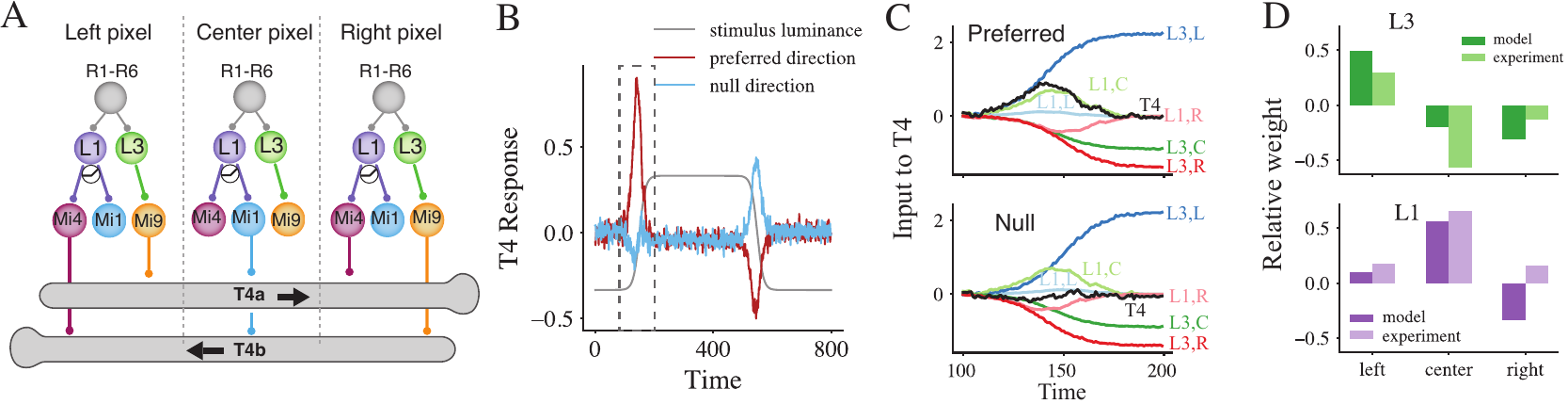}
\vspace{-2mm}
  \caption{Experimentally observed and learned motion-detection networks exhibit similar response properties and synaptic connectivity. (A) The \textit{Drosophila} ON motion-detection pathway~\cite{takemura2017comprehensive,borst2023flies}. For comparison with the model (Fig.~\ref{fig:fig0}), we simplify the circuit by estimating the effective weights of inputs from L1 and L3 to T4 directly, as the intermediate neurons are thought to perform primarily contrast normalization~\cite{matulis2020heterogeneous}. (B) The unrectified response of a T4 analog in the trained ReSU network to a moving grating mirrors experimental results: the strongest, sharply peaked response occurs for an advancing bright (“ON”) edge in the preferred direction, whereas weaker responses to “OFF” edges in the null direction can be suppressed by thresholding. (C) Synaptically weighted contributions of each first-layer output channel to the T4 response (black) for ON-edge motion in the preferred and null directions around stimulus onset (dashed box in B). (D) Second-layer synaptic weights in the trained model reproduce the majority of sign patterns and approximate the relative amplitudes of synaptic inputs onto T4a in \textit{Drosophila}~\cite{takemura2017comprehensive}. Because experimental weights are based on synapse counts without knowledge of neuronal gain factors, we compare L1 and L3 inputs separately. 
 }
  \label{fig:fig4}
  \vspace{-2mm}
\end{figure*}

\section{A Two-Layer ReSU Network Trained on Natural Stimuli Detects Motion Similar to \textit{Drosophila}}


In this section, we train the second layer of the ReSU network on the outputs of the L1 and L3 analogs from adjacent pixels (Figure \ref{fig:fig0}). We find that, post-training, the ReSU network's neuronal responses and synaptic weights align with the empirical results in \textit{Drosophila} (Figure \ref{fig:fig4}).

To motivate the architecture of the model network (Figure \ref{fig:fig0}), we summarize a few biological facts about \textit{Drosophila} visual processing. ON-motion signals are computed separately from OFF-motion signals by processing the rectified outputs of L1 and the outputs of L3 neurons \cite{takemura2013visual,borst2023flies}. We simplify the \textit{Drosophila} circuit (Figure \ref{fig:fig4}A) by directly connecting the outputs of L1 and L3 neurons to the T4 neuron (Figure \ref{fig:fig0}) because the intermediate neurons are thought to primarily implement contrast normalization \cite{matulis2020heterogeneous}. 
Whereas each L1 and L3 neuron is stimulated by a single ``pixel'' \cite{rivera2014wiring}, the second layer integrates outputs of L1 and L3 corresponding to several adjacent pixels \cite{takemura2017comprehensive}. Assuming a one-dimensional retina in the model, we consider three consecutive pixels.

To train the ReSU network, we present the scalar contrast time series, $\{x_t\}$, obtained by scanning the rows of panoramic natural images. By performing CCA on the (centered) past, $\mathbf{p}_t^1$, and future, $\mathbf{f}_t^1$, lag vectors we learn temporal filters and compute the output of L1 and L3 neurons in each pixel, $\mathbf{z}_t: z_1(t) = \Psi_1 \mathbf{p}_t^1, z_2(t) = [\Psi_2 \mathbf{p}_t^1]_+$. 

A ReSU in the second layer receives the outputs of L1 and L3 neurons from three adjacent pixels---six channels total---$\mathbf{y}_t = [z_1^L(t), z_2^L(t),z_1^C(t), z_2^C(t), z_1^R(t), z_2^R(t)]^\top \in \mathbb{R}^6$, where the superscripts $L, C, R$ stand for left, center and right pixels, respectively. We learn the second-layer spatial filters by performing CCA on (uncentered) $\mathbf{p}_t^2 = \mathbf{y}_t$ and $\mathbf{f}_t^2=\mathbf{y}_{t+\text{lag}}$ (Eq. \ref{eq:CCA2}) with $\mathbf{y}_t$ elicited by the same contrast time series, $\{x_t\}$, with a motion-induced temporal offset between adjacent pixels. Note that, after rectification in L1, its output is no longer uncorrelated with the output of L3. For details, see Supplementary Material.

We next evaluate the response of the trained two-layer ReSU network to moving grating stimuli (ON/OFF; Figure~\ref{fig:fig4}). Filtering the first-layer output with the $i=1$ ReSU of the second layer produced a motion-direction-independent response, whereas unrectified filtering with the $i=2$ ReSU yielded a stronger response in the preferred than in the null direction (Figure~\ref{fig:fig4}B). The residual null-direction response can be eliminated through appropriate thresholding within the ReSU. In a biological setting, neurons could acquire
$i=2$-like filters if the $i=1$ projection were suppressed, for example via inhibitory interneurons. The two T4 subtypes (T4a and T4b) can develop opposite direction selectivity when trained on stimuli translating in opposite directions (Figure~\ref{fig:fig4}). Given the known gating of synaptic plasticity by postsynaptic activity, we further speculate that when training stimuli contain motion in both directions, such selectivity could emerge through plasticity gated by the rectified output of the corresponding T4 neuron.

To better understand the underlying mechanism of motion selectivity, we plot the input to a T4 ReSU, specifically, the L1 and L3 outputs of three adjacent pixels weighted by the spatial filter (Figure \ref{fig:fig4}C). The differential response of T4 to preferred-direction motion is driven by the left L3 input ramping up prior to the drop of the right and center L3 input and, to a lesser degree, center L1 input. After taking into account the simplification of the model architecture compared to the actual connectome, this mechanism aligns with the experiment, Figure 5C,D in \cite{borst2023flies}. 

Finally, we compare the spatial canonical directions of the first-layer outputs with the feedforward synaptic weights onto T4a neurons from connectomics  and find qualitative similarity (Figure~\ref{fig:fig4}D).

\section{Discussion}

We propose self-supervised ReSU networks as an alternative to the conventional supervised ReLU networks both for modeling biological circuits and for constructing artificial neural networks. A two-layer ReSU network captures salient features of the motion-detection pathway in \textit{Drosophila}. Stacking ReSU layers opens a path to construct deep networks that learn progressively more complex features.

We use 1D motion detection as a well-established, tractable—but nontrivial—testbed to demonstrate that self-supervised learning on natural stimuli can yield bio-plausible circuits. Since 2D motion perception builds on a pair of horizontal and vertical 1D detectors, our framework can be readily generated to model 2D motion detection.

While our focus here is on the \textit{Drosophila} visual pathway, ReSU networks can be applied to other sensory modalities and species. Our work was originally motivated by graded-potential neurons whose outputs are rectified by downstream synapses—a property shared by both invertebrate and vertebrate retinas, as well as the \textit{C. elegans} nervous system. Nonetheless, ReSUs can also approximate spiking neurons under a firing-rate representation. 

In the current formulation, the temporal memory and prediction horizon are specified \textit{a priori}, rather than learned by the neuron. Future work will aim to develop methods for automatically determining the optimal memory length based on the input statistics.

We demonstrated a self-supervised learning of non-trivial features in a two-layer ReSU network, but whether this approach generalizes to deeper networks remains an open question. We have recently extended this work to a three-layer ReSU network better capturing the fly motion detection pathway~\cite{icassp}, Figure \ref{fig:fig4}A. While the current two-layer network was trained on images translated in only one direction (left to right), the three-layer ReSU network learns from bidirectionally translated images further establishing a foundation for learning more complex hierarchical features in deeper architectures. 

While the theoretically derived neuronal temporal filters qualitatively resemble those observed experimentally using calcium imaging, further validation is needed. Testing the theory would require measurements of neuronal responses and temporal filters at higher temporal resolution, for instance using voltage indicators. Additional promising approaches include examining adaptation to stimulus statistics and probing circuit mechanisms through targeted manipulations such as neuron ablation or silencing.

As early sensory processing is largely feedforward, we expect that ReSU networks trained on natural stimuli offer a principled and interpretable model. Hierarchical features learned by self-supervised ReSU networks should support diverse behavioral tasks. As one dives deeper into the brain, the contribution of top-down feedback---and the number of feedback loops---progressively increases. In the future, it would be interesting to explore how to incorporate top-down feedback in our current framework. One potential avenue is to combine the theoretical foundations of the data-driven ReSU framework~\cite{pughe2025neurons} with the data-driven controller neuron model~\cite{moore2024neuron}.

We call our algorithm potentially local because future work will aim to extend the neural network of static units with local learning rules which performs CCA between two concurrent data streams \cite{lipshutz2021biologically} to our setting of past-future CCA. We anticipate that incorporating temporal structure will not pose significant challenges as it can be implemented locally within each neuron through the use of distinct ion channels with different time constants. Another difference of our framework from the CCA formulation of \cite{lipshutz2021biologically} is that the neuronal output corresponds to a projection of only past inputs onto the canonical direction, with future inputs used exclusively for learning. A similar self-supervised problem has been solved  using local learning rules in a static setting \cite{golkar2020biologically}, and we therefore do not expect it to present major difficulties.

\section{Acknowledgments} 
We are grateful to R. Behnia, T. Clandinin, D. Clark, C. Karaneen, I. Nemenman, S.E. Palmer, E. Schneidman, E. Schomburg, D. Schwab, A. Sharafeldin, M. Silies, and N. Srebro for helpful discussions, to R. M. Haret and T. Gollisch for sharing their data. Some of this work was initiated and performed at the Aspen Center for Physics, which is supported by NSF grant PHY-2210452.

\bibliography{refs}



\end{document}


\maketitle

\renewcommand{\thefigure}{S\arabic{figure}}

\section{Derivation of past-future CCA for OU processes}

Optimal choices of latent variables are not always those that minimize prediction error in an $L^2$ sense. Indeed, from an information theoretic perspective, it is motivated to capture components of the system state that are most correlated with future dynamics \cite{chechik2005}. Canonical correlation analysis (CCA) both accomplishes this task \textit{and} reduces it to a spectral problem, rendering its solution simple to compute.

Consider lag vectors ${\bf p}_t$ and ${\bf f}_t$, representing past and future intervals of the system dynamics centered at time $t$. In one dimension, CCA seeks a direction ${\bf v}\in\text{span}\set{{\bf p}_t}$ and ${\bf u}\in\text{span}\set{{\bf f}_t}$ maximizing 
$$\text{corr}\left({\bf u}^\top {\bf f}_t, {\bf v}^\top {\bf p}_t\right).$$

That is, CCA determines an observation of the local past, ${\bf v}^\top{\bf p}_t$, that is maximally correlated with an associated observation of the future, ${\bf u}^\top{\bf f}_t$. If ${\bf v}^\top{\bf p}_t$ is large at time $t$, then ${\bf u}^\top{\bf f}_t$ is expected to be large in the coming moments, representing an \textit{IF-THEN} type observation of the external world. 
When determining an $r$-dimensional compression of the system state, CCA seeks a set of $r$ vectors 
$$\Psi =\begin{bmatrix}
    {\bf v}_1^\top \\ {\bf v}_2^\top \\ \vdots \\ {\bf v}_r^\top
\end{bmatrix} \in \R^{r \times m}$$
that maximize the correlation of ${\bf p}_t$ and ${\bf f}_t$ within their span. 

This can be formulated mathematically by the following optimization 
problem, which is derived in \cite{arun1990balanced}, and which reproduces (5) in the main text:
\begin{equation}\label{cca_def2}
  \max_{\mathbf{\Psi}\in\mathbb{R}^{r\times m}}
    \bigl\lVert \mathbb{E}\!\bigl[\mathbf{C}_{ff}^{-1/2}\,\mathbf{f}_t(\mathbf{\Psi}\mathbf{p}_t)^\top\bigr]\bigr\rVert_{F},
    \quad
    \text{s.t.}\;
    \mathbf{\Psi}\mathbf{C}_{pp}\mathbf{\Psi}^\top= \mathbf{I}_r.
\end{equation}
Here, $\mathbf{f}_t, \mathbf{p}_t$ are the stochastic past and future lag vectors of 
length $m$, and 
${\bf C}_{ff} = \mathbb{E}[\mathbf{f}_t\mathbf{f}_t^\top], {\bf C}_{fp} = \mathbb{E}[\mathbf{f}_t\mathbf{p}_t^\top], {\bf C}_{pp} = \mathbb{E}[\mathbf{p}_t\mathbf{p}_t^\top]$.
We observe that 
\begin{equation}\label{cca_eq2}
\begin{aligned}
  \mathbb{E}\!\bigl[\mathbf{C}_{ff}^{-1/2}\,\mathbf{f}_t(\mathbf{\Psi}\mathbf{p}_t)^\top\bigr] 
&  =   \mathbf{C}_{ff}^{-1/2}\,\mathbb{E}\!\bigl[ \mathbf{f}_t \mathbf{p}_t^{\top} \bigr] \mathbf{\Psi}^\top \\
&  =   \mathbf{C}_{ff}^{-1/2}\, \mathbf{C}_{fp} \mathbf{\Psi}^\top.
\end{aligned}
\end{equation}
Applying the change of variables $\tpsi =\mathbf{\Psi}\mathbf{C}_{pp}^{1/2}$
to \eqref{cca_eq2} and to the constraint of \eqref{cca_def2}, we have 
\begin{equation}\label{cca_eq3}
\begin{aligned}
  \mathbb{E}\!\bigl[\mathbf{C}_{ff}^{-1/2}\,\mathbf{f}_t(\mathbf{\Psi}\mathbf{p}_t)^\top\bigr] 
&  =   \mathbf{C}_{ff}^{-1/2}\, \mathbf{C}_{fp} \mathbf{C}_{pp}^{-1/ 2} \tpsi^\top
\quad \text{and} \quad 
 \mathbf{\Psi}\mathbf{C}_{pp}\mathbf{\Psi}^\top= \tpsi \tpsi^\top.
\end{aligned}
\end{equation}
Substituting \eqref{cca_eq3} into \eqref{cca_def2} we find that the solution 
to optimization \eqref{cca_def2} 
is equal to
\begin{equation}\label{cca_eq4}
  \max_{\tpsi \in\mathbb{R}^{r\times m}}
    \bigl\lVert \mathbb{E}\!\bigl[ \mathbf{C}_{ff}^{-1/2}\, \mathbf{C}_{fp} \mathbf{C}_{pp}^{-1 / 2} \tpsi^\top  \bigr] \bigr\rVert_{F},
    \quad
    \text{s.t.}\;
    \tpsi  \tpsi^\top = \mathbf{I}_r.
\end{equation}
The solution to the reformulated optimization problem \eqref{cca_eq4} is 
obtained by the singular value decomposition (SVD) $\mathbf{C}_{ff}^{-1/2}\, \mathbf{C}_{fp} \mathbf{C}_{pp}^{-1 / 2} = \mathbf{U \Sigma V}^\top$  \cite{blum2020foundations}.
Assuming the singular values on the diagonal of $\mathbf{\Sigma}$ are ordered from 
largest to smallest, the optimum of \eqref{cca_eq4} is achieved by the Frobenius norm of the 
upper-left $r \times r$ submatrix of
$\mathbf{\Sigma}$. The $\tpsi$ that achieves that maximum is the matrix of the 
first $r$ columns of $\mathbf{V}$, which we denote $\mathbf{V}_r$.  

Furthermore, since $\mathbf{V}_r$ achieves the maximum of \eqref{cca_eq4}, using 
the change of variables $\tpsi =\mathbf{\Psi}\mathbf{C}_{pp}^{1/2}$, the maximum 
of \eqref{cca_def2} is achieved by $\mathbf{\Psi}=\mathbf{V}^\top \mathbf{C}_{pp}^{-1/2}$.

\subsection*{Derivation of mutual information between $\mathbf{z}_t$ and $\mbf$}
In this section we prove that 
the mutual information between $\mathbf{z}_t$ and $\mbf_t$ is 
equation (8) in the main text:
\begin{align}
I_r = -\frac{1}{2} \sum_{i=1}^r \log(1 - \sigma_i^2).
\end{align}
We start by observing that the covariance matrix of $[\mathbf{z}_t \, \mbf_t]$ is 
the $(r + m) \times (r + m)$ matrix 
\begin{align}
\mathbf{\Sigma}_{\mathbf{z}_t, \mbf_t} = 
\begin{bmatrix}
\mathbf{I}_r & \mathbf{\Sigma}^\top \mathbf{U}^\top \mathbf{C}_{ff}^{1/2} \\
\mathbf{C}_{ff}^{1/2} \mathbf{U} \mathbf{\Sigma} & \mathbf{C}_{ff}
\end{bmatrix}.
\end{align}
The off-diagonal blocks are obtained via the equations
\begin{align}
\mathbb{E} [\mbf_t \mathbf{z}_t^\top]  & = \mathbb{E}[\mbf_t (\mpsi \mbp_t)^{\top}] \\
& = \mathbb{E}[\mbf_t \mbp_t^{\top}] \mpsi^\top.
\end{align}
Using the definition of $\mpsi$ and that $\mathbf{C}_{ff}^{-1/2}\, \mathbf{C}_{fp} \mathbf{C}_{pp}^{-1 / 2} = \mathbf{U \Sigma V}^\top$, we have
\begin{equation}
\begin{aligned}
\mathbb{E} [\mbf_t \mathbf{z}_t^\top]  & = \mathbf{C}_{fp} \mathbf{C}_{pp}^{-1/2} \mathbf{V} \\
& = \mathbf{C}_{ff}^{1/2} \mathbf{C}_{ff}^{-1/2} \mathbf{C}_{fp} \mathbf{C}_{pp}^{-1/2} \mathbf{V} \\ 
& = \mathbf{C}_{ff}^{1/2} \mathbf{U} \mathbf{\Sigma} \mathbf{V}^\top \mathbf{V} \\ 
& = \mathbf{C}_{ff}^{1/2} \mathbf{U} \mathbf{\Sigma}.
\end{aligned}
\end{equation}
Using the formula for the mutual information between Gaussian vectors \cite{cover2006elements}, 
we obtain 
\begin{align}\label{mutual_inf_gauss}
I_r = \frac{1}{2} \log \bigg( \frac{| \mathbf{I}_r | |\mathbf{C}_{ff}|}{| \mathbf{\Sigma}_{\mathbf{z}_t, \mbf_t} |} \bigg)
\end{align}
where $|\cdot | $ denotes the matrix determinant. 
Substituting the formula for the determinant of a $2 \times 2$ block matrix \cite{horn2012matrix}, 
we have
\begin{align}
| \mathbf{\Sigma}_{\mathbf{z}_t, \mbf_t} | & = | \mathbf{I}_r | | \mbc_{ff} - (\mbc_{ff}^{1/2} \mathbf{U \Sigma) (\Sigma}^\top \mathbf{U}^\top \mbc_{ff}^{1/2}) |.
\end{align}
Applying the determinant property $| AB| = |A| |B|$ along with the fact that 
orthogonal matrices have determinant of one, we have 
\begin{equation}\label{matrix_det}
\begin{aligned}
| \mathbf{\Sigma}_{\mathbf{z}_t, \mbf_t} | & = | \mbc_{ff}^{1/2} (\mathbf{I}_r - \mathbf{U \Sigma \Sigma}^\top \mathbf{U}^\top ) \mbc_{ff}^{1/2}| \\
& = | \mbc_{ff} | |(\mathbf{I}_r - \mathbf{U \Sigma \Sigma}^\top \mathbf{U}^\top )| \\
& = | \mbc_{ff} | |\mathbf{U} (\mathbf{I}_r - \mathbf{\Sigma \Sigma}^\top) \mathbf{U}^\top| \\
& = | \mbc_{ff} | |(\mathbf{I}_r - \mathbf{\Sigma \Sigma}^\top)|.
\end{aligned}
\end{equation}
Applying equation \eqref{matrix_det} to \eqref{mutual_inf_gauss}, we obtain 
\begin{align}
I_r = \frac{1}{2} \log\bigg(\frac{1}{| \mathbf{I}_r - \mathbf{\Sigma \Sigma}^\top |}  \bigg) = 
-\frac{1}{2} \sum_{i=1}^r \log(1 - \sigma_i^2)
\end{align}
where $\sigma_i = \mathbf{\Sigma}_{i, i}$.

\section{Gaussian processes}
In Section 2 of the main text, we observe that CCA can be extended to time-invariant 
Gaussian processes (GPs) with integrable covariance kernel as long as the 
past vector is sufficiently long. Here, we numerically 
test the necessary length $\ell$ of the past lag vector  
$\bm p_t$ that incorporates enough information that CCA is approximately 
optimal in the sense of equation (5). 

We start by constructing a grid $t_j = \Delta_t j$ with $\Delta_t = 0.05$ for 
$i = 1,\dots, 2 \ell$.\footnote{The conclusions of this section do not depend on the choice of $\Delta_t$. The lag vector lengths discussed here are only relevant for a particular choice of $\Delta_t$. If $\Delta_t$ is multiplied by a constant, conclusions hold when dividing the lag by the same constant.} We then we construct
the $2\ell \times 2\ell$ GP covariance matrix
\begin{align}
\bm \Sigma = \begin{bmatrix}
\mathbf{C}_{pp} & \mathbf{C}_{pf} \\
\mathbf{C}_{fp} & \mathbf{C}_{ff}
\end{bmatrix}
\end{align}
where $\bm \Sigma_{ij} = k(t_i - t_j)$. Here, $k$ is  with rational quadratic covariance kernel
\begin{align}\label{eq:gp_rq}
k(t - t') = \left(1+ \frac{|t-t'|^2}{2\alpha l^2}\right)^{-\alpha}
\end{align}
with $l = \alpha = 1$ used in the main text. 
We compute the canonical directions $\bm \Psi$ and 
the canonical correlations $\sigma_1,\dots,\sigma_{\ell}$ as described in Section 1 of the main text. 
We show that for sufficiently large $\ell$, increasing $\ell$ adds negligible 
information about the future. That is, for large $\ell$, increasing $\ell$ further
results in adding 
near-zero components to canonical directions $\bm \Psi$ and also 
results in adding zero canonical correlation coefficients.
Specifically, we compare the canonical correlations and directions for lag $\ell$ 
and compare them to those computed for lag $2\ell$. 
Throughout this section, we set the past and future lags to both be of the same 
size, though fixing the size of the future lag vector does not change the results.

\begin{figure}[ht]
\centering
\includegraphics[width=0.8\linewidth]     {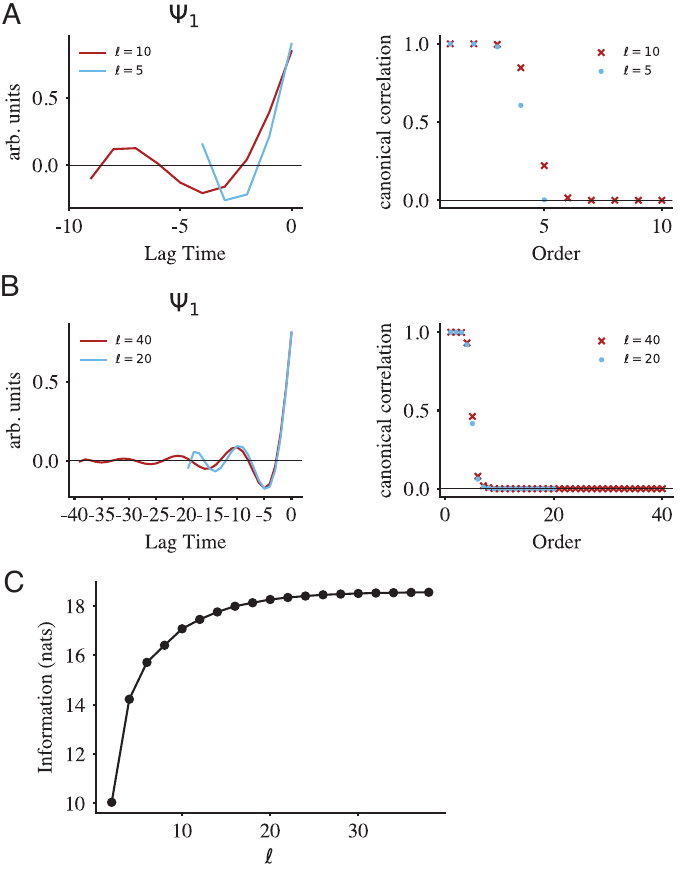}
\caption{ 
CCA for a GP with rational quadratic kernel converges as 
the lag length, $\ell$, increases. We demonstrate the convergence of 
temporal filters (or canonical directions) $\mpsi_1$
in the top row and in the second row we plot the corresponding correlation coefficients.
(A) For $\ell = 5$ and $\ell=10$, there is still substantial disagreement 
in the fourth and fifth canonical directions and in $\mpsi_1$. 
(B) For $\ell = 20$, the filter $\mpsi_1$ and the canonical directions 
already closely approximate the $\ell = 40$ case.
(C) The mutual information between $\bm p_t$ and $\bm f_t$ converges 
as we increase the lag length $\ell$.
}
\label{fig:figS1}
\end{figure}

In Fig. \ref{fig:figS1}A, we compare $\ell = 5$ and $\ell=10$. 
Even a lag of $\ell = 5$ captures significant information about the future, but there is 
substantial information lost compared to $\ell = 10$ (see the difference between 
the fourth canonical correlations). 
We compare $\ell = 20$ and $\ell = 40$ in Fig. \ref{fig:figS1}B. With a lag of $\ell = 20$, 
CCA is well-approximated by the CCA for $\ell = 40$. 
Furthermore, the number of canonical correlations needed for capturing nearly 
all information about the future is contained in five canonical directions. We plot
convergence of the first canonical direction $\mpsi_1$. The convergence of the next 
canonical directions is similar to that of the first canonical direction. 

We also compute the mutual information of past and future lag 
vectors $\bm p_t, \bm f_t$ for various $\ell$
using equation (8). 
The results in Fig. \ref{fig:figS1}C demonstrate that 
the mutual information rapidly approaches a limit. This reinforces 
that a lag of around $\ell = 20$ captures nearly all information 
about the future. 

The results of this section, unlike in the main text, 
compute CCA with access to the exact covariance matrix 
in contrast to a covariance matrix approximated from data. 
We observe numerically that with more observation noise,
there are fewer canonical correlations near one. Intuitively, 
the more noisy the past and future observations are, the less
correlated the past will be with the future.

For many commonly-used GP covariance kernels (e.g., squared exponential,
rational quadratic, Mat\'ern, etc.), the cross-covariance 
matrix $\mathbf{C}_{fp}$ is numerically low-rank. That is, the singular values 
of $\mathbf{C}_{fp}$ decay quickly. This phenomenon is a result of the 
the fact  that covariance kernels tend to be smooth (in the sense of frequency
content) away from zero and is discussed at length 
in \cite{ambikasaran2016gps}.
The matrix $\mathbf{C}_{fp}$ being numerically low-rank does not guarantee
that the matrix $ \mathbf{C}_{ff}^{-1/2}\, \mathbf{C}_{fp} \mathbf{C}_{pp}^{-\top / 2}$
is also numerically low-rank. 
However, we observe empirically that this tends to be the case 
and equivalently that past/future CCA for GPs tends to produce few 
canonical correlations near one. We demonstrate this property of
the rational quadratic kernel in the second column of Fig \ref{fig:figS1}.

\section{Details of numerical simulations}
\subsection{Contrast from natural scenes}

We used a natural scene dataset from  \cite{meyer2014panoramic},  which can be downloaded from \href{https://pub.uni-bielefeld.de/record/2689637}{here}.  This dataset consists of 421 panoramic luminance-calibrated naturalistic images of natural scenes. Each image contains 927×251 pixels and subtends 360° horizontally and 97.4° vertically. 

To simulate the optical system and photoreceptor transduction of \textit{Drosophila}, the grayscale luminance images were converted into contrast images with a blurring step and contrast computation step as outlined in \cite{chen2019asymmetric,zavatone-veth2020}. Briefly, the pixel values of original pictures, denoted as $I$, were blurred with a two-dimensional Gaussian filter $f_{\text{blur}(x,y)}$
\begin{align}
    I_{\text{blur}} &= I*f_{\text{blur}}(x,y),\\
    f_{\text{blur}}(u, v) &= \frac{1}{2\pi\lambda_{\text{blur}}^2} \exp(-\frac{u^2 + v^2}{2\lambda_{\text{blur}}^2}), \quad u_- \leq u \leq u_+, \quad v_- \leq v \leq v_+.
\end{align}
The filter extends to $\pm 3 \lambda_{\text{blur}}$, i.e., $u_+ = v_+ = |u_\_| = |v_\_|$, where $\lambda_{\text{blur}} = \frac{5.3°}{2\sqrt{2 \ln 2}}$ is chosen to mimic the spatial resolution of the fly’s ommatidia, and $*$ denotes convolution. 

Then, the blurred images were converted to contrast signals
\begin{align}
    c(x,y) &= \frac{I_{\text{blur}} - I_{\text{mean}}(x,y)}{I_{\text{mean}}(x,y)},\\
    I_{\text{mean}}(x,y) &= I_{\text{blur}}*f_{\text{local-mean}}
\end{align}
where$f_{\text{local-mean}}$ is a 2-dimensional Gaussian filter similar to that of $f_{\text{blur}}$ but with a larger spatial scale than $\lambda_{\text{blur}}$ \cite{chen2019asymmetric}. Scalar contrast time series was generated by scanning a selected row of the contrast image with constant velocity. A Matlab code convert raw pixel pictures to blurred contrast is available at: https://github.com/ClarkLabCode/SynapticModel/tree/master.

\subsection{Comparing the response to staircase stimuli}
In fly vision study, staircase luminance/contrast stimuli are widely used to probe the temporal properties of early visual neurons. Since our model accounts for the response of neurons a synapse behind photoreceptors, the L1/L2/L3 neurons, the luminance was logarithmically transformed to approximate the responses of photoreceptors \cite{laughlin1989role}.

The staircase stimulus was generated by concatenating up and down transients. Each upward transient is generated by hyperbolic function $\tanh(x)/2 + b_0$ with $x$ uniformly taking value from -10 to 10 and the constant base $b_0$ runs from 1 to 3 indicating the elevated plateaus. The downward transients are mirror of the upward transients.  A small obervation noise was also introduced ($\sigma = 0.01$), the second column of figure 4 shows the average predicted responses from our model.

\subsection{Training a two-layer ReSU network on natural scene}
To demonstrate that a two layer ResU network can perform non-trivial computation. We trained the network with contrast profiles from natural scene. The input to the first layer ReSU are lag vectors $\mathbf{p}_t,\mathbf{f}_t$ of scalar image contrast $x_t$ as described in the previous subsection. In our numerical simulation, both the memory and horizon are set to be 50 and the observation noise level is 0.05. The output is a 2-D vector time series $\mathbf{z}_t$, the first component is the projection of input lag vector $\mathbf{p}_t$ on the first filter $\Psi_1$, and the second component is the projection onto the second filter with rectification, i.e. $z_2(t) = [\Psi_2 \mathbf{p}_t]_+$. The input to the second layer network consist of output of three adjacent ``channels'' of first layer
$\mathbf{y}_t = [z_1^L(t), z_2^L(t),z_1^C(t), z_2^C(t), z_1^R(t), z_2^R(t)]^\top \in \mathbb{R}^6$, where the superscripts $L, C, R$ stand for left, center and right channel, respectively. This set up mimic the spatial pooling of downstream motion sensitive neurons such as T4 and T5. In practice, these three channels are separated by about 5 degree which corresponds 13 horizontal pixel on the panoramic pictures. Thus, the time series of $z_{1,2}^C(t), z_{1,2}^R$ are time-shifted versions of $z_{1,2}^L$. We also normalize each channel by its standard deviation and introduced small observation noise. Finally, we obtain the past and future lag vectors for the second layer
$\mathbf{p}_t^2  = \mathbf{y}_t, \mathbf{f}_t^2 = \mathbf{y}_{t+\text{lag}}$, the covariance matrices are defined as
$\mathbf{C}_{fp} = \mathbb{E}(\mathbf{f}_t^2 \mathbf{p}_t^{2\top})$, $\mathbf{C}_{ff} = \mathbb{E}(\mathbf{f}_t^2 \mathbf{f}_t^{2\top})$, $\mathbf{C}_{pp} = \mathbb{E}(\mathbf{p}_t^2 \mathbf{p}_t^{2\top})$, and the objective function of for the second layer is the same as Equation 6 in the main text. For the numerical results shown in the figure 5, the lag length is set to be 5.

After applying CCA on the lag vectors of the second layer, we obtained two filters for the second layer. The output of the network is then projection of $\mathbf{p}_t$ onto the second filter $\Psi_2^2$. The weights of $\Psi_2^2$ are then compared with the synaptic weights between LMCs and T4 neurons in fly as shown in figure 5 in the main text \cite{takemura2017comprehensive}.

\subsection{Adaptation of filters to change of input statistics}

\begin{figure}[ht]
\centering
\includegraphics[width=0.6\linewidth]     {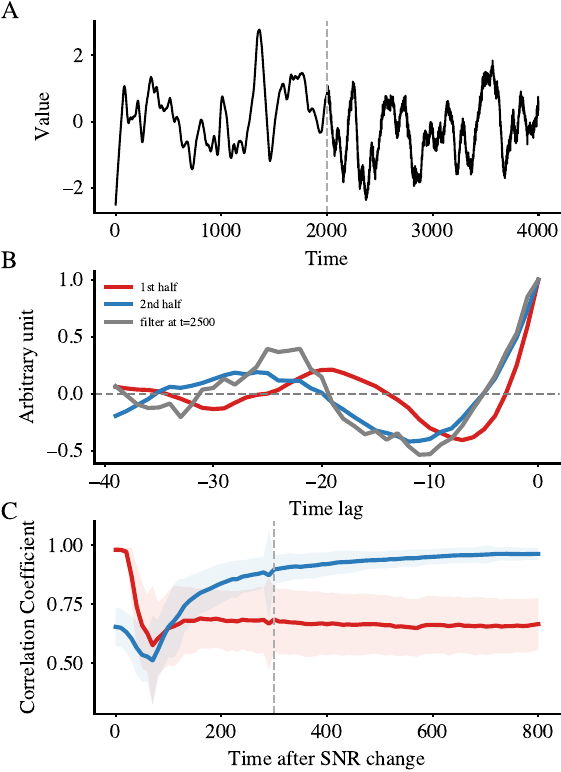}
\caption{ 
Filter in ReSU adapts to the input statistics quickly. (A) A input generated from GP with a rational quadratic kernel. It changes noise level at $t = 2000$. (B) The second filter obtained from offline CCA for the first half (red) and the second half (blue) trajectory. The filter at $t= 2500$ is estimated using all the data up to $t=2500$ but with an exponential discount. (C) Correlation coefficient of the ``current'' filter with that of the reference filters (as shown in B). Shaded region indicates standard deviation from 100 independent simulations. Parameters: $m=40, h = 20$. 
}
\label{fig:figS2}
\end{figure}

We demonstrate that filters in the ReSU network adapt rapidly to changes in input statistics. We used time series generated from a Gaussian process with a rational quadratic kernel (equation \eqref{eq:gp_rq}) where $\alpha = \ell = 1$. The time series had two distinct regimes: the first half had high signal-to-noise ratio (SNR, $\sigma= 0.01$) and the second half had low SNR ($\sigma = 0.1$), as shown in Figure \ref{fig:figS2}A.

To quantify the adaptation speed when input statistics changed at $t = 2000$, we tracked how the ``current filter'' correlated with two reference filters: one estimated from the first half of the trajectory (high SNR) and one from the second half (low SNR). We estimated the current filter at each time point by computing a past-future covariance matrix from preceding input data, where data points were weighted by an exponential decay function $e^{-|t' -t|/\tau}$. This weighting scheme gives greater importance to recent observations, with $\tau = 400$ in our simulations, where $t$ is the current time point and $t'$ represents previous time points.

When the input SNR changed, the filter's correlation with the first reference filter decreased while its correlation with the second reference filter increased rapidly. The filter adapted to the low-SNR statistics in approximately 300 time steps, less than 10 times the memory lag (Figure \ref{fig:figS2}C).

\subsection{Example of the OU process}
The example of OU process shown in Figure 2 was simulated by the following system
\begin{align}
    X_{t+1} &= AX_t + B\xi_t,\\
    y_t &= CX_t + \eta_t,
\end{align}
where
\begin{equation}
    A = 
    \begin{bmatrix}
    0.6&0.6&0\\
    -0.6&0.6&0\\0&0&0.4
    \end{bmatrix}, \quad
    B = 
    \begin{bmatrix}
        0.17,-0.15,0.28
    \end{bmatrix}, \quad
    C = 
    \begin{bmatrix}
        0.78, 0.53,1
    \end{bmatrix}
\end{equation}
and $\xi_t \sim \mathcal{N}(0,1), \eta_t \sim \mathcal{N}(0,0.1)$.

\subsection{Experimental data}
The temporal filters of retinal ganglion cells shown in figure 3 were obtained from \cite{liu2015spike} and the dataset can be downloaded \href{https://datadryad.org/dataset/doi:10.5061/dryad.7r7n7#usage}{here}. The synaptic weights of the ON motion detection pathway shown in figure 5 was obtained from figure 6 of \cite{takemura2017comprehensive}.

\subsection{Hyperparameters}
We list all the hyper parameters used to learn the temporal filters in the simulations, such as the past and future lag, $m, h$, and the observation noise level, $\sigma$ in table \ref{param}.

\begin{table}[h]
  \caption{parameters}
  \label{param}
  \centering
  \begin{tabular}{lcccccccc}
    \toprule
    parameter & \multicolumn{3}{c}{Fig. 2}&     Fig. 3 & Fig. 4 & Fig. 5  & Fig. S1  & Fig. S2        \\
    \cmidrule(r){2-4}
    & B     & C & D     \\
    \midrule
    $m$       & 25 & 75  & 50   & 75        & 50      & 50 & & 40  \\
    $h$       & 25 & 25  & 50   & 25        & 50     & 50& & 20\\
    $\sigma$  & 0  & 0.1 & 0.05 &  0.01, 0.4& 0.5, 0.05 &0.05, 0.01 & 0 & 0.01, 0.1\\
    \bottomrule
  \end{tabular}
\end{table}

In Fig. 4, when testing the action of filters on the staircase input, a small observation noise $\sigma = 0.002$ was added. Similarly, in Fig. 5B, a small observation noise ($\sigma  = 0.005$) was added to the step-like contrast.

\subsection{Code availability}
All the code in this project are available at a GitHub repository: \url{https://github.com/ShawnQin/ReSU.git}

\printbibliography